\documentclass[aps,prb,preprint,groupedaddress,showpacs,floatfix]{revtex4-1}
\usepackage{amsmath,amssymb,latexsym,epic,eepic,epsfig,graphicx,psfrag}
\usepackage[latin1]{inputenc}

\newcommand{\fig}[1]{Fig.~\ref{#1}}

\usepackage{bm}
\usepackage{comment}

\begin{document}

% Use the \preprint command to place your local institutional report
% number in the upper righthand corner of the title page in preprint mode.
% Multiple \preprint commands are allowed.
% Use the 'preprintnumbers' class option to override journal defaults
% to display numbers if necessary
%\preprint{}

%Title of paper
\title{Atomic-scale model for the contact resistance of the
  nickel-graphene interface}

% repeat the \author .. \affiliation  etc. as needed
% \email, \thanks, \homepage, \altaffiliation all apply to the current
% author. Explanatory text should go in the []'s, actual e-mail
% address or url should go in the {}'s for \email and \homepage.
% Please use the appropriate macro foreach each type of information

% \affiliation command applies to all authors since the last
% \affiliation command. The \affiliation command should follow the
% other information
% \affiliation can be followed by \email, \homepage, \thanks as well.
\author{Kurt Stokbro}
\email[]{kurt.stokbro@quantumwise.com}
\author{Mads Engelund}
\author{Anders Blom}
\affiliation{QuantumWise A/S,\\
 Lers{\o} Parkall\'{e} 107,
  DK-2100 Copenhagen, Denmark}
\homepage{http://quantumwise.com}

%\email[]{Your e-mail address}
%\homepage[]{Your web page}
%\thanks{}
%\altaffiliation{}
%Collaboration name if desired (requires use of superscriptaddress
%option in \documentclass). \noaffiliation is required (may also be
%used with the \author command).
%\collaboration can be followed by \email, \homepage, \thanks as well.
%\collaboration{}
%\noaffiliation

\date{\today}

\begin{abstract}
  We perform first-principles  calculations of electron transport
  across a nickel-graphene interface. Four different geometries are
  considered, where the contact area, graphene and nickel surface
  orientations and the
  passivation of the terminating graphene edge are varied. We find covalent bond
  formation between the graphene layer and the nickel surface, in
  agreement with other theoretical studies. We calculate the
  energy-dependent electron transmission for the four systems and find that
  the systems have very similar edge contact resistance,
  independent of the contact area between nickel and graphene, and in
  excellent agreement with
  recent experimental data. A simple model where  graphene is bonded with a
  metal surface shows that the results are generic for covalently
  bonded graphene, and the minimum attainable edge contact
  resistance is twice the ideal edge quantum contact resistance of graphene.
\end{abstract}

% insert suggested PACS numbers in braces on next line
%\pacs{73.40.-c, 73.63.-b, 72.10.-d, 72.80.Vp}
% insert suggested keywords - APS authors don't need to do this
%\keywords{}

%\maketitle must follow title, authors, abstract, \pacs, and \keywords
\maketitle
\section{\label{sec:intro}Introduction}
Recently there has been an increasing interest in the use of
graphene for electronic devices. One of the  outstanding
questions is the magnitude of the contact resistance between graphene and metal
electrodes, since a high contact resistance will limit the performance
of field-effect transistors~\cite{Xia2010}. There have been several experimental
investigations of the contact resistance of the metal-graphene
interface using four- or two-probe
measurements~\cite{Nagashio2010a, Venugopal2010, Blake2009, Russo2010, Malec2011} and the transfer length
method~\cite{Venugopal2010, Xia2011}, however, currently there is no
clear consensus on the value and the dependence of the
contact resistance on contact area, temperature and applied gate
potential. Thus, there is a need for complementary theoretical studies
which can give insight about the physical mechanism at play at the
metal-graphene interface.

Previous first-principles theoretical studies have focused on the effect of charge
transfer between metal and graphene on the contact
resistance~\cite{Khomyakov2009,Lopez2010,Maassen2010, Ran2009}. In this
paper we add new knowledge to the understanding of the graphene-metal
contact by investigating the effect of covalent bond formation on the
contact resistance.  We will present quantum transport calculations of the electron
transfer from a free suspended graphene sheet to a nickel contact  through different
metal-graphene contact geometries, where we vary the orientation of
the graphene and the contact area between nickel and graphene. Graphene forms a strong
covalent bond with nickel~\cite{Khomyakov2009} which is similar to the bond
formation between graphene and cobalt, palladium and titanium, thus, the theoretical
predictions will also be relevant for these systems. We find that the contact resistance is  independent of the orientation of
the graphene, as well as of the contact area to the metal,
in excellent quantitative agreement with recent
experimental observations~\cite{Nagashio2010a}.

\section{The calculations}
\begin{figure}[tbp]
\begin{center}
  \includegraphics[width=\linewidth]{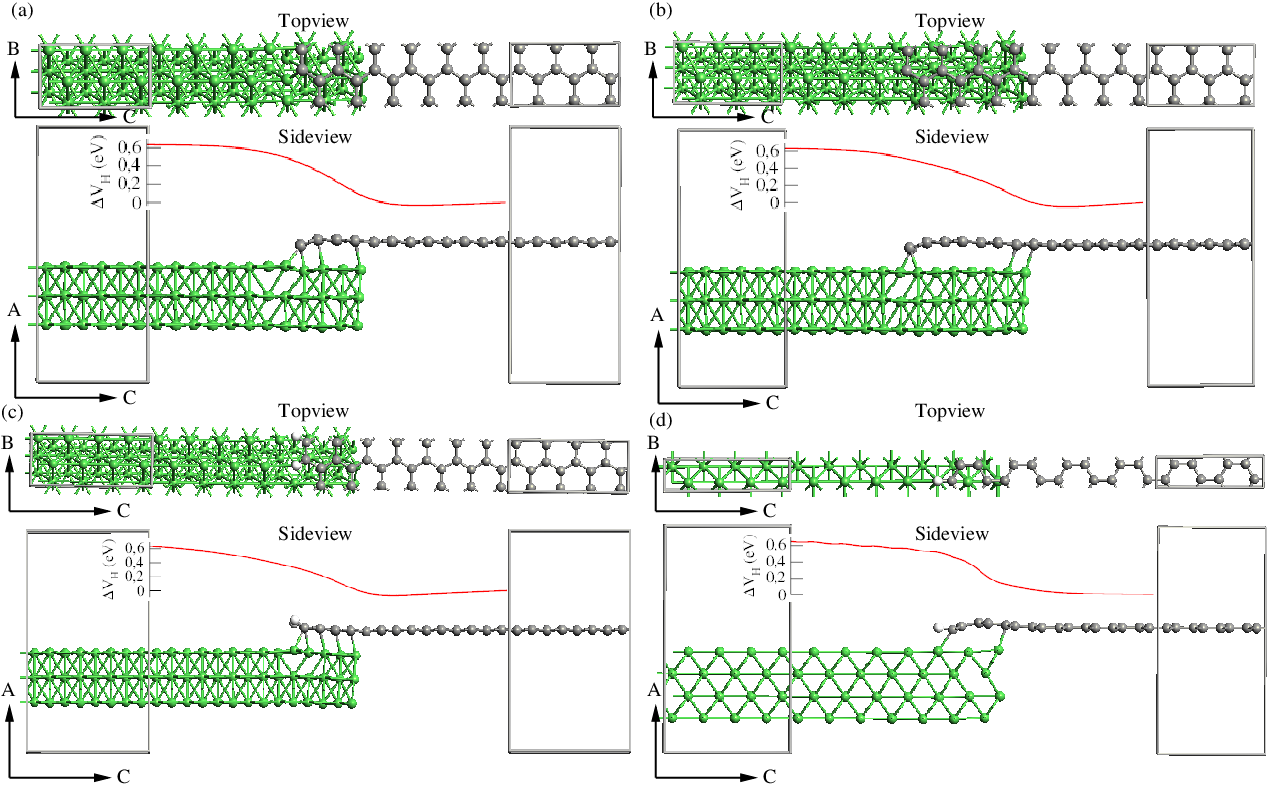}
\end{center}
  \caption{Top view (B-C plane) and side view (A-C plane) of the four systems investigated in
    this paper. The transport direction is along the C direction.
    (a) Zigzag edge graphene on top of a Ni(111) surface with 4~\AA\
    binding overlap. (b) Similar to (a) but with 8~\AA\ overlap. (c)
    Similar to (a) but with a hydrogen-passivated terminal edge. (d) Armchair
    edge graphene on top of a Ni(100) surface with 4~\AA\ overlap,
    and hydrogen-passivated terminal edge. The inset (red curve) in
    each figure shows the average electrostatic potential in
    the vacuum region along the C
    direction. The potential has been averaged over the B direction for a
    fixed  A coordinate ($A$=18~\AA). }
  \label{fig:structures}
\end{figure}

\fig{fig:structures} illustrates the four different
graphene-nickel interfaces considered in this paper. For systems (a),
(b), and (c) the graphene is adsorbed
on a Ni(111) surface and
oriented with a zigzag edge in the transverse transport
direction (direction B in \fig{fig:structures}). In system (d) it is
adsorbed on a Ni(100) surface and has an armchair edge in the transverse transport
direction. With these choice of orientation the
lattice mismatch between nickel and graphene is about 1\%. In order to
simplify the comparison between the different systems, we fix the
lattice constant of the graphene, and strain the nickel surface by 1\% to
obtain a commensurate supercell for both systems.

The overlap region between nickel and graphene is 4~\AA\ in (a), (c), (d),
while it is 8~\AA\ in (b). In (c), (d) the graphene edge is passivated by
hydrogen. Thus, the systems represent very different types of
graphene-nickel interfaces.

For the calculations we have used
Atomistix ToolKit (ATK)~\cite{ATK12.2}, which is a density-functional
theory code using numerical localized atomic basis sets. ATK
allows for simulating open systems through use of a non-equilibrium Green's
function (NEGF) formalism as described in
Ref.~\onlinecite{Brandbyge-2002-PRB}. The systems in \fig{fig:structures} are
heterogeneous along the C direction (the left and right electrodes are
not the same) and thus even at zero bias the system is not periodic in the
transport direction. When calculating
the electrostatic potential we therefore employ a Poisson solver which combines the FFT method
in the A and B directions (in which the structure is periodic) with a multigrid solver for
the C direction~\cite{openMx}, where Dirichlet boundary conditions are
used for the open system.

We used a double-$\zeta$ polarized
basis set for expanding the electronic density. This basis set
consists of  15 basis orbitals for each nickel atom, with 2 sets of orbitals
of {\it s}-type, 1 of {\it p}-type and 2 of {\it d}-type.
The nickel basis functions had an extended range compared to
the default ATK basis set values, in order to obtain a good description of
the nickel work function. The radii of the basis functions were
4.46~\AA, 4.46~\AA,  and 2.84~\AA\ for the s, p and d channels, respectively. For
each carbon atom 13 orbitals per atom were used, with 2 sets of orbitals
of {\it s}-type, 2 of {\it p}-type and 1 of {\it d}-type. The
cut-off radius of the orbitals were 2.39~\AA,  2.86~\AA\ and 2.86~\AA, for the
s, p and d channels, respectively. Other technical parameters were
a density mesh cut-off of 150 Rydberg
and 9 k-points in the B-direction where the
structures are periodic.

For the exchange-correlation we used
the Perdew-Zunger parametrized  local spin density
approximation (LSDA)~\cite{Perdew1981} since it has been demonstrated
to give excellent results for the geometry of the nickel-graphene
interface\cite{Thygesen-2011}. To determine the geometry of the
interface, we first optimized the
relative distance between the nickel surface and the graphene layer, with otherwise fixed atom
positions. We find a distance of 2.00~\AA, in good agreement with
Ref.~\onlinecite{Khomyakov2009}. Subsequently we relaxed all atoms in the interface region
such that the force on each atom was less than 0.05~eV/\AA.

The relaxed structures are illustrated in \fig{fig:structures}. There is a covalent
bond formation between the graphene and nickel atoms. The bond
formation destroys the $\pi$-conjugation of the graphene sheet and it
is no longer flat, but buckled with   distances
between the nickel surface and the graphene sheet  in the range 1.85--2.3~\AA.

\fig{fig:structures} also shows the electrostatic profile along the
C-direction in the vacuum
region. We see that the vacuum level is 0.6~eV higher above the nickel
surface, compared to the graphene layer, corresponding to a 0.6~eV
larger work function $W$ of nickel compared to graphene. This is in
excellent agreement with the difference in the measured  work function
of nickel
$W_{100}$=5.22~eV, $W_{111}$=5.35~eV~\cite{Baker-1971}, and graphene, $W$=4.6~eV~\cite{Oshima-1997}.

\begin{figure}
  \centering
  \includegraphics[width=\linewidth]{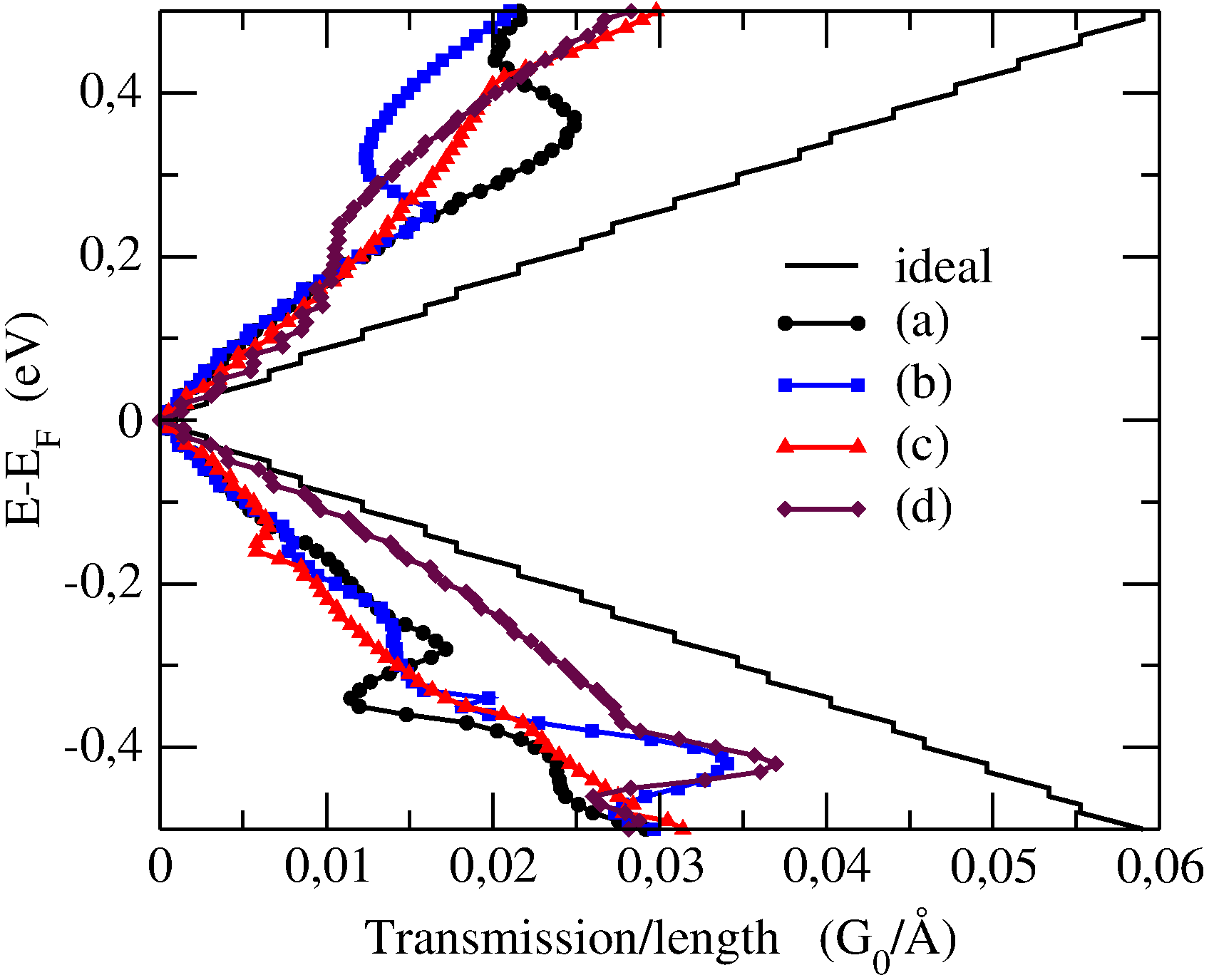}
  \caption{Transmission coefficient per transverse line segment at zero bias for
    the four different systems illustrated in \fig{fig:structures}. Also shown is the
    transmission coefficient of an ideal graphene layer. }
  \label{fig:transmission}
\end{figure}

We next calculate the transmission coefficient for each geometry and
the result is illustrated in \fig{fig:transmission}. We note that the
transmission coefficients  for the two spin channels are very similar, and the
figure therefore only  shows the total transmission of both spin channels. For the
transmission calculation we used 501 k-points in the B direction. This corresponds
to an equivalent graphene ribbon width of 2134~\AA~\cite{Lopez2010}.
In the energy range $[-0.1, 0.1]$~eV the transmission coefficients are almost
identical, and have a V-shaped form with a slope 0.06~$G_0/\mathrm{\AA \, eV}$.
\fig{fig:transmission} also shows the transmission of an ideal
graphene sheet
calculated with the same parameters. Also in this case the transmission spectrum has the form of
a wedge with a singularity at the Fermi level, this time
with a slope of 0.12~$G_0/\mathrm{\AA \, eV}$.

Thus, all investigated graphene-nickel systems have a similar contact resistance, which is a factor $\sim$2
 larger than the ideal quantum contact resistance of a graphene
 sheet. In the following we will analyze the calculations to
 understand the origin of the similar contact resistance of the four systems.

\begin{figure}[tbp]
\begin{center}
  \includegraphics[width=\linewidth]{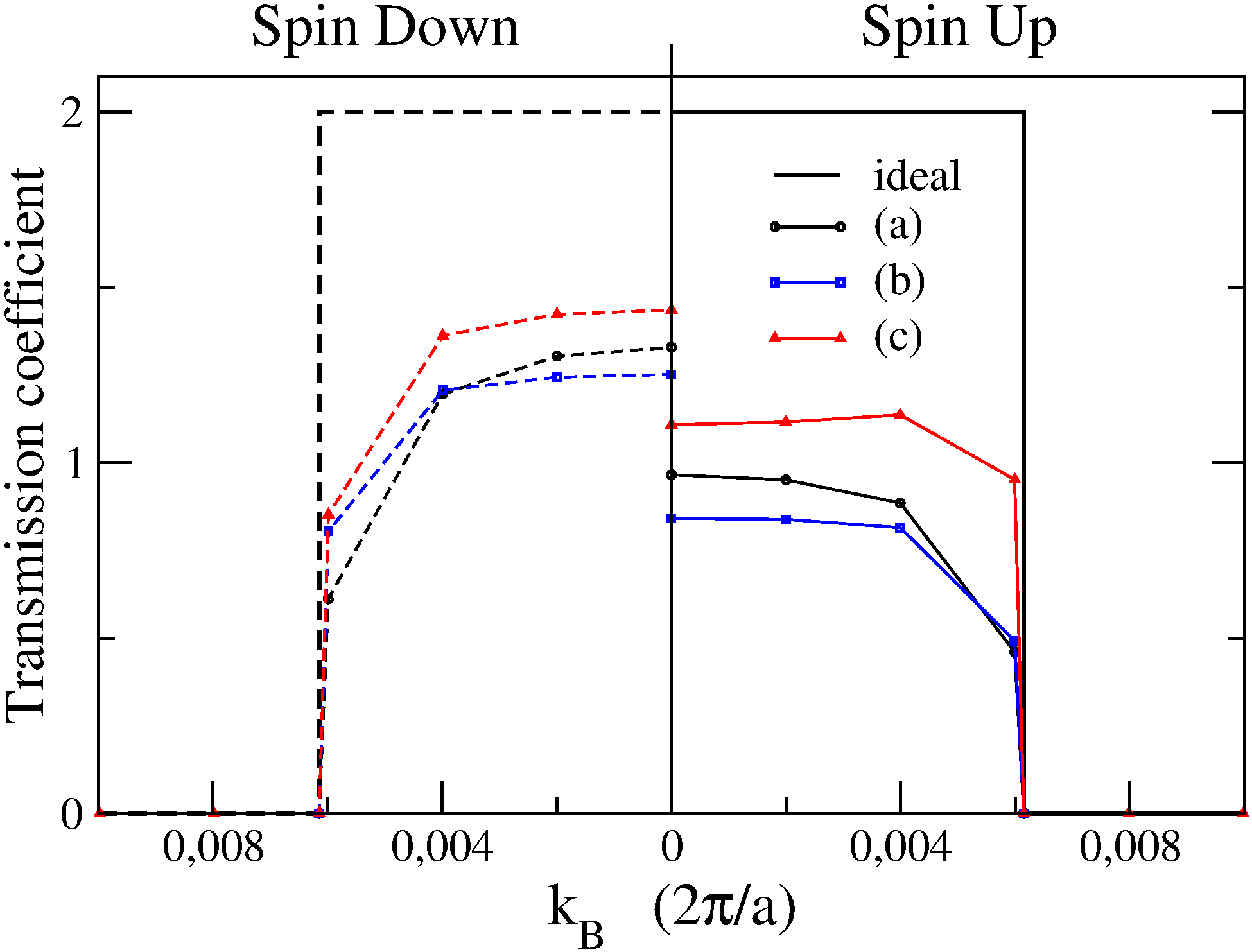}
\end{center}
  \caption{Transmission coefficient at $E = 0.05$~eV as function of the
    transverse k-point in the B-direction, $k_B$. The right-hand part of the graph shows the spin-up
    component while the left-hand part shows the spin down
    component. An ideal graphene sheet has two transmission channels for
   $k_B < 0.0062\times2\pi/a$.  }
  \label{fig:transkp}
\end{figure}

The transmission coefficient in \fig{fig:transmission} is obtained by
averaging the transmission coefficient over the k-points in the B
direction, $k_B$. In \fig{fig:transkp} we show how the transmission
coefficient varies as function of $k_B$ at the energy $E-E_F = 0.05$~eV.
For the perfect graphene sheet there are two transmission channels
for $k_B < 0.0062\times 2\pi/a$. We see that for system (a), (b), and (c),
approximately half of the
channels transmit through the interface, with the total transmission
coefficient for the two channels varying in the range 0.8--1.4. Thus, the systems behave qualitatively
similar, but there are quantitative differences. It is interesting to
note that system (b), which has a larger bonding area than system (a),
has a slightly smaller transmission coefficient. Thus, the bonding
area does not seem to be an important factor.

\begin{figure}[tbp]
\begin{center}
  \includegraphics[width=\linewidth]{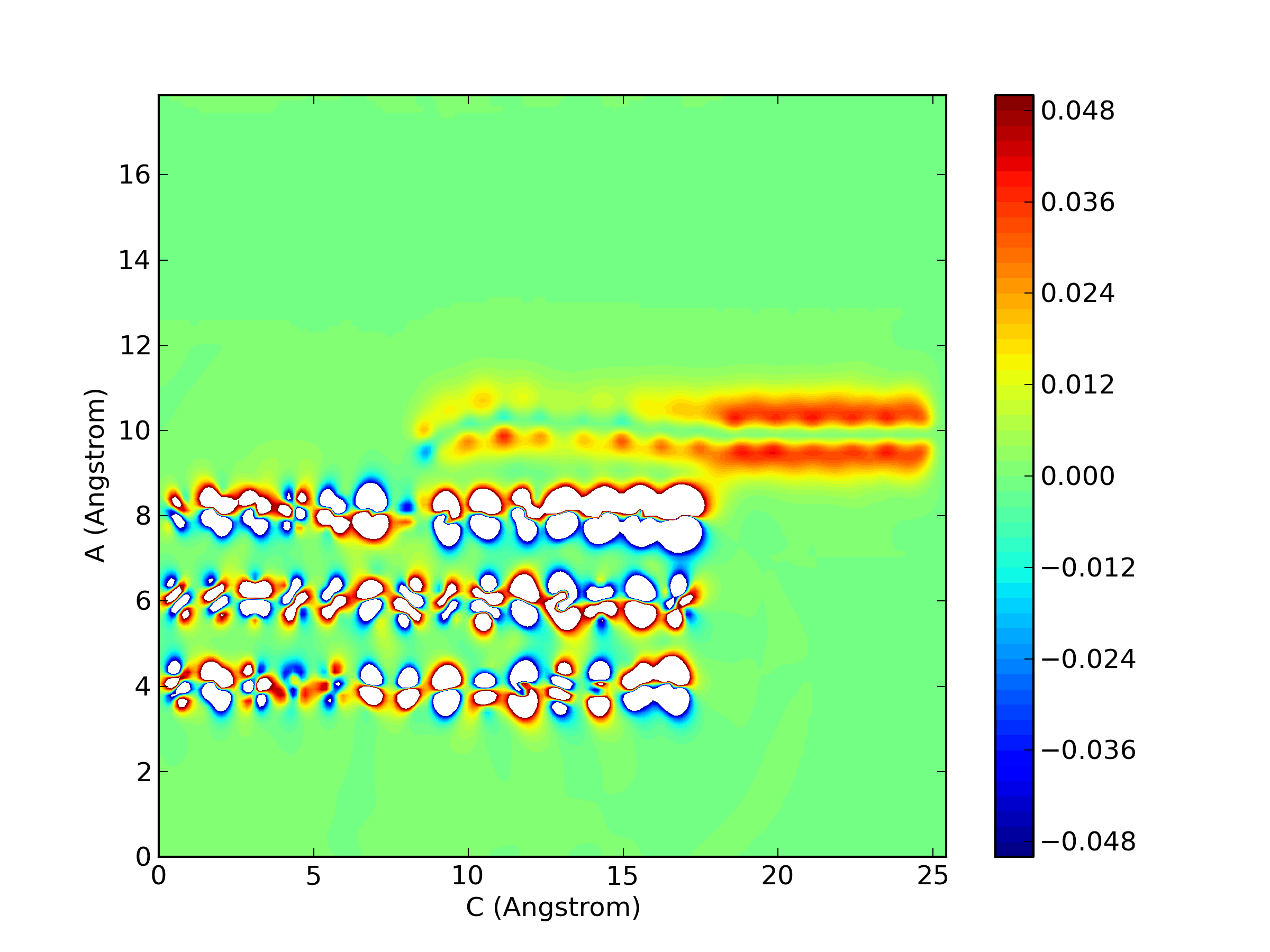}
\end{center}
  \caption{Contour plot of the C-component of the linear response current density in the A-C plane
    for states with energy 0.05~eV, averaged in the B-direction (arbitrary units). }
  \label{fig:current_density}
\end{figure}

To gain further insight into the transport mechanisms, we have also
calculated the  current density in system (b)
(without inclusion of non-local potential corrections~\cite{Li2008}),
from the states with energy 0.05~eV.  The result is shown in
\fig{fig:current_density} and is a real-space view of the current
density of the states giving rise to the curve (b) in \fig{fig:transkp}.

The figure illustrates how the
current incident from the right (from the graphene side) gets
transmitted through the device. The current density in the graphene
layer shows a symmetry corresponding to carbon $\pi$-electrons
carrying the current.  The current density in the graphene sheet drops
at the boundary between the graphene atoms
bonded to the nickel surface and the non-bonded graphene atoms. This
means that the main resistance occurs at the interface between
non-bonded and bonded graphene atoms, which explains why the
bonding area between nickel and graphene is not important.

\begin{figure}[tbp]
\begin{center}
  \includegraphics[width=\linewidth]{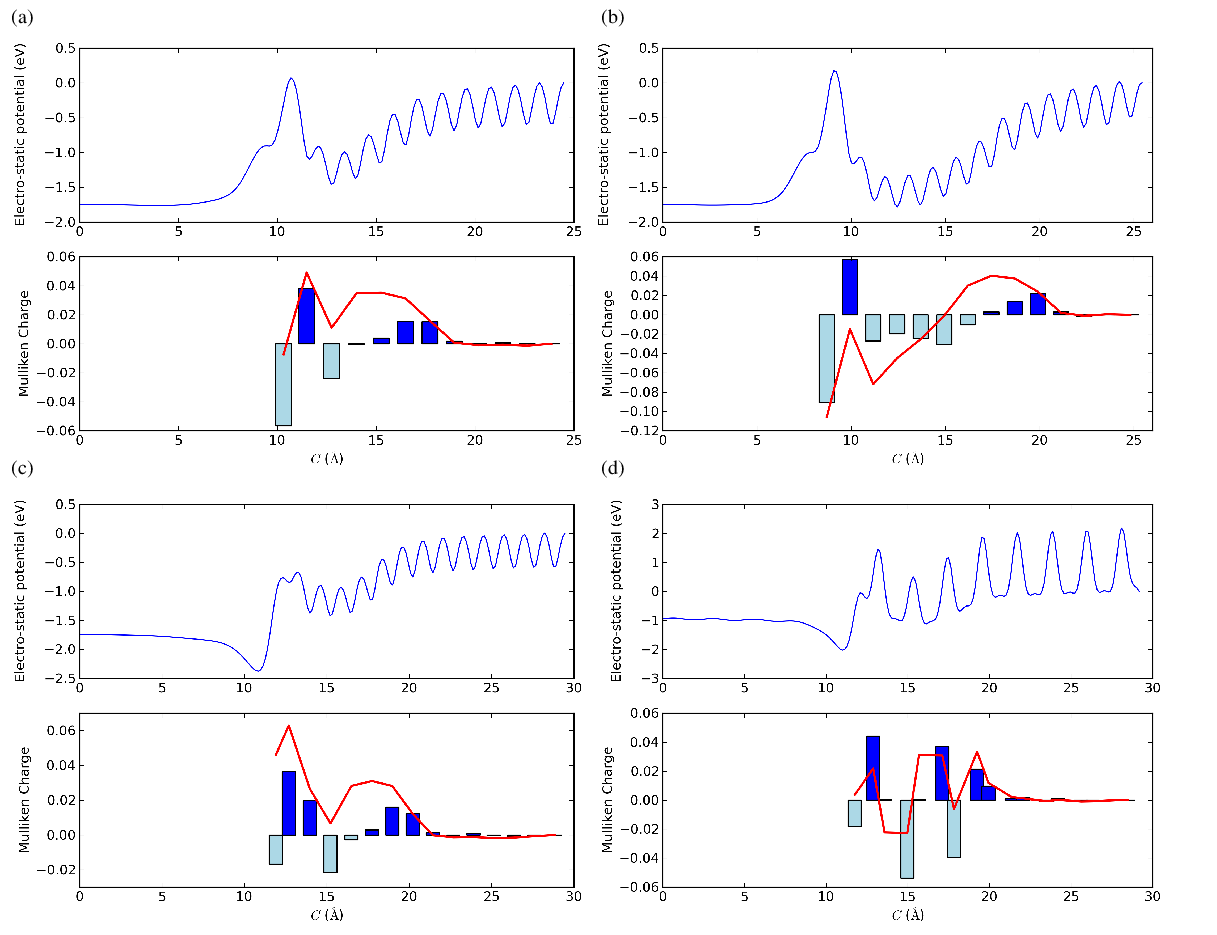}
\end{center}
  \caption{Average electrostatic difference potential and Mulliken
    charge  inside the graphene sheet plotted along the C
    direction. Results are shown for each of the four systems
     in \fig{fig:structures}. The electrostatic difference potential is shown for
     fixed  A=10~\AA\ (corresponding to  the plane of the
    graphene sheet) and  is averaged over the
    B-direction. The zero-point of the potential is defined as the
    potential at the  right-hand edge of the
    cell. The bar chart shows the Mulliken charge on each atom in
  the graphene sheet, and the full line shows the accumulated charge
  from free-hanging graphene to the edge atoms, i.e.\ the charge is
  accumulated from right to left.}
  \label{fig:potential}
\end{figure}

\fig{fig:potential} shows the electrostatic difference potential
within the plane of the graphene sheet for the four systems. The  plot should be compared with
the average electrostatic potential in the vacuum region, illustrated in \fig{fig:structures};
note that in \fig{fig:structures} the profiles correspond to a plane far away from the graphene sheet (A=18~\AA),
whereas in \fig{fig:potential} we cut right through the graphene.  The figure
also shows the corresponding Mulliken charges $-e(m-z_v)$, where  $m$ is the valence Mulliken population of
each atom, and  $z_v$ is
the valence charge. The solid (red) line shows the accumulated charge in the
graphene layer.

For all systems we find that close to the nickel edge, for $C>$15~\AA,  there is electron transfer
from graphene to nickel, as a result of the  0.6~eV higher work function
of the nickel surface compared with graphene. This charge transfer
gives rise to a lowering of the electrostatic potential in the
graphene sheet. At the edge of the graphene sheet, 10~\AA $< C <$ 15~\AA,
the amount of charge transfer depends on the edge termination. For the
non-terminated surfaces, (a) and (b), there is an electron accumulation at
the edge, corresponding to a negative edge charge. This gives rise to
a positive jump in the electrostatic potential. For the H-terminated
surfaces, (c) and (d), there is an electron depletion at the edge, thus
a positive edge charge and a downwards jump in the
electrostatic potential.

From this we may conclude that there is no
relation between the  contact resistance and the charge transfer
between nickel and graphene, in contrast with weakly bonded systems
where charge transfer has been observed to play an import role~\cite{Maassen2010}.

\section{A model system}
To illustrate that the observed transmission coefficient is rather
generic for covalently bonded graphene, we   have set up a
simple model system consisting of an aluminum surface
and a graphene layer. The system is not relaxed, and the
transmission coefficient is calculated using  an
extended H\"{u}ckel model~\cite{Stokbro-2010}. The model in
Ref.~\onlinecite{Stokbro-2010}  allows for self-consistently adjusting the
onsite elements, but in order to have the most simple model this
option is not used in the current study.

\begin{figure}[tbp]
\begin{center}
  \includegraphics[width=\linewidth]{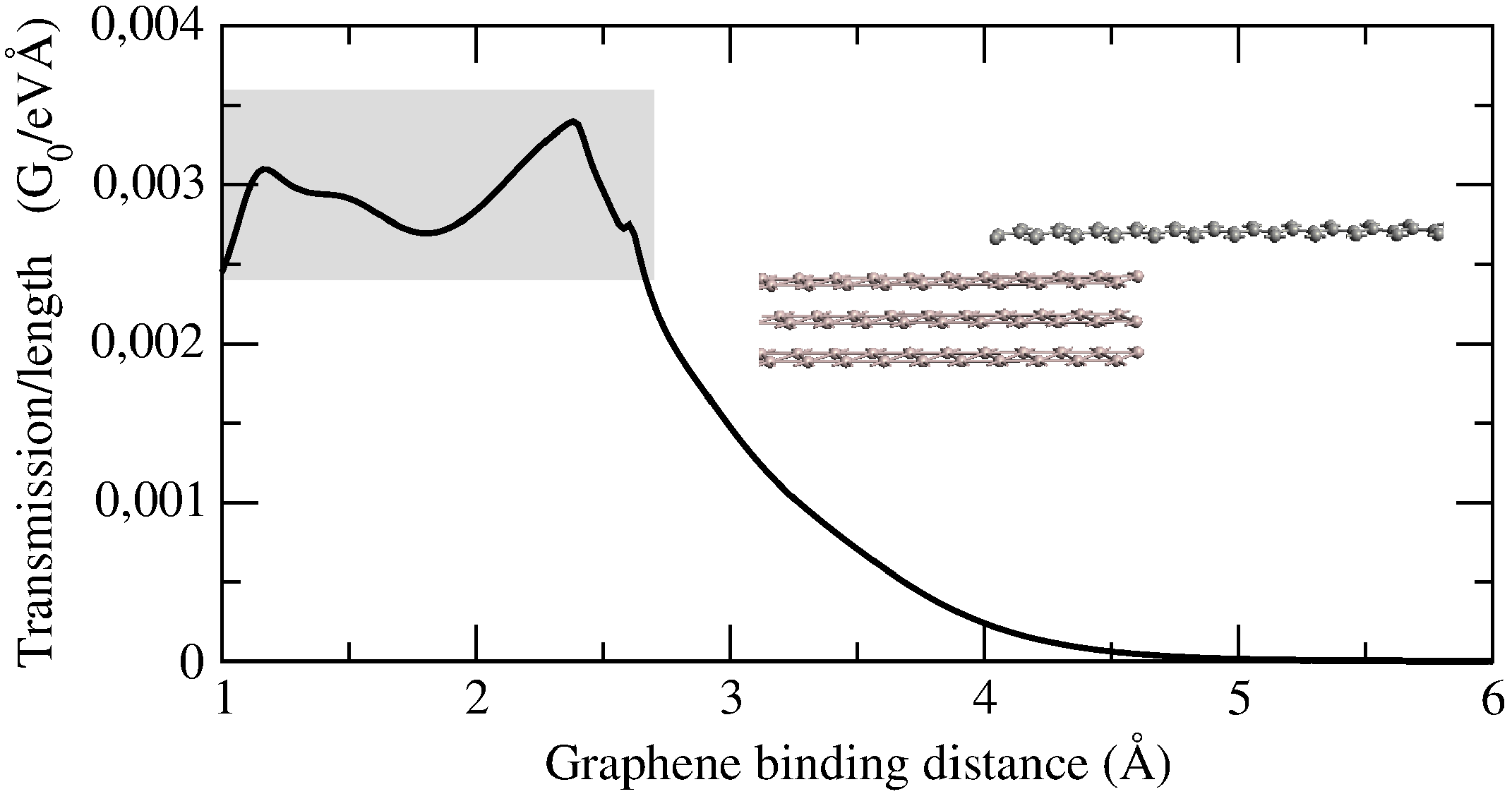}
\end{center}
  \caption{Transmission coefficient per transverse line segment at energy
    $E-E_F=0.05$~eV as function of the graphene-surface adsorption distance. Calculations
    are for a model system consisting of a graphene sheet
    in contact with a surface made
    of aluminum. The gray area illustrates the variations of  the transmission coefficient per transverse line segment at energy
    $E-E_F=0.05$~eV for  system (a), (b), (c) and (d)  in
    \fig{fig:transmission}.}
  \label{fig:al-graphene}
\end{figure}

\fig{fig:al-graphene} shows the average transmission coefficient  per transverse line segment
at the energy $E-E_F=0.05$~eV  as function of the  distance between
the graphene overlayer and the surface. By varying the adsorption height we change the
effective interaction between the surface and the graphene. The gray
area illustrates the   variations in the transmission
coefficients at this energy for the four systems in
\fig{fig:transmission}. In the range 1 \AA $< d < $2.5
\AA there is a strong interaction between graphene and the metal
surface, and we see that in this range the transmission coefficient is
similar to the system (a), (b), (c), (d)  in
    \fig{fig:transmission}.

Based on these results, we suggest the following model for the electron transmission for
a covalently bonded graphene-metal system. The system can be divided into two parts: (i)
the non-bonded graphene,  and (ii) the metal surface with the
covalently bonded
graphene. We may now diagonalize system (ii) into left- and
right-going modes. When graphene is strongly bonded to the metal
surface, we may regard graphene as an extension of the metal surface
and there will be equally many left and right going modes in the
graphene layer.  An incoming left-going electron  from system (i)
may couple with either the left- or right-going modes in system
(ii). In the strong coupling regime, the carbon atoms in system (ii) will
be enough perturbed by the metal surface that both  left- and right-going modes there
bear very little resemblance to the modes in the non-bonded graphene. Thus, the incoming
electrons from system (i)  will on average have  the same
coupling strength with left- and right-going modes in system (ii),
and thus approximately half of the incoming current is transmitted through the
system, as the results in \fig{fig:transmission} show.

\section{Discussion}
Recent experiments on the contact resistance of the nickel-graphene
interface~\cite{Nagashio2010a} found that the contact resistance did
not depend of the contact area, and has  an edge contact
resistance of  $\sim$$800\ \mathrm{\Omega \mu m}$ at room
temperature. Our calculations also show that the contact resistance is
independent of the contact area. From the
transmission spectra in \fig{fig:transmission}, and the approximation
$T(E) \approx 0.06\ (\mathrm{eV \AA})^{-1} |E-E_F|$   the
edge contact resistance can be calculated from
\begin{equation}
1/R = G_0 \times0.06\ (\mathrm{eV \AA})^{-1}  \int |E-E_F|
\frac{e^{(E-E_F)/k_B T}}{(1+e^{(E-E_F)/k_B T})^2}\frac{ dE}{k_B T}.
\end{equation}
 Using
a room-temperature Fermi distribution in the electrodes, we obtain an edge
contact resistance of $\sim$$600\ \mathrm{\Omega \mu m}$. This is in excellent
accordance with the experimentally observed value, and shows that the
contact resistance in the experiment arises from the ballistic  quantum contact resistance.

In summary, we have presented calculations demonstrating that the contact resistance of a
nickel-graphene junction is independent of the contact area and the
direction of the graphene sheet. The edge
contact resistance is $\sim$$600\ \mathrm{\Omega \mu m}$,
corresponding to twice the ideal quantum
contact resistance of pure graphene and in excellent agreement with
experimental data. Additional model calculations predict that this
result is generic for strongly bonded graphene on metal surfaces.

%\begin{acknowledgments}\label{sec:acknowledgements}

%\end{acknowledgments}
\bibliography{paper}

\end{document}